\documentclass[aip, jcp, reprint]{revtex4-1}
\pdfoutput=1
\usepackage[version=3]{mhchem}
\usepackage{xcolor}
\usepackage[perpage,symbol]{footmisc}
\usepackage{tikz}
\usepackage{pgfplots}
\usetikzlibrary{shapes.geometric, arrows, positioning}
\usepackage{subcaption}
\usepackage{dcolumn}
\usepackage{multirow}

\usepackage{hyperref}

\hypersetup{
    colorlinks=true,
    linkcolor=black,
    citecolor=black,
    filecolor=black,
    urlcolor=black,
}

\newcommand\kpoint{$\mathbf{k}$-point}
\newcommand\kpoints{$\mathbf{k}$-points}

\newcommand\CZTS{Cu$_2$ZnSnS$_4$}

\newcommand*{\diff}[2]{\frac{\text{d}{#1}}{\text{d}{#2}}}

\begin{document}

\title
{Crystal structure optimisation using an auxiliary equation of state}

\author{Adam J. Jackson}
\author{Jonathan M. Skelton}
\author{Christopher H. Hendon}
\author{Keith T. Butler}
\affiliation{Centre for Sustainable Chemical Technologies and Department of Chemistry, University of Bath, Claverton Down, Bath, BA2 7AY, United Kingdom}
\author{Aron Walsh}
\email[]{a.walsh@bath.ac.uk}
\affiliation{Centre for Sustainable Chemical Technologies and Department of Chemistry, University of Bath, Claverton Down, Bath, BA2 7AY, United Kingdom}
\affiliation{Global E$^3$ Institute and Department of Materials Science and Engineering, Yonsei University, Seoul 120-749, Korea}


\begin{abstract}
  Standard procedures for local crystal-structure optimisation involve numerous energy and force calculations. It is common to calculate an energy--volume curve, fitting an equation of state around the equilibrium cell volume. This is a computationally intensive process, in particular for low-symmetry crystal structures where each isochoric optimisation involves energy minimisation over many degrees of freedom.
  Such procedures can be prohibitive for non-local exchange-correlation functionals or other `beyond' density functional theory electronic structure techniques, particularly where analytical gradients are not available. We present a simple approach for efficient optimisation of crystal structures based on a known equation of state.
  The equilibrium volume can be predicted from one single-point calculation, and refined with successive calculations if required. The approach is validated for PbS, PbTe, ZnS and ZnTe using nine density functionals, and applied to the quaternary semiconductor \ce{Cu2ZnSnS4} and the magnetic metal-organic framework HKUST-1.
\end{abstract}

\maketitle


\section{Introduction}
The standard operating procedure for computational investigations in solid-state chemistry is to begin with a crystal structure --  obtained either from diffraction studies or through chemical analogy -- and to optimise the lattice shape, volume and internal positions to minimise all forces. 
It is from this equilibrium crystal structure (athermal for the majority of electronic-structure approaches) that the full range of material response functions (e.g. elastic, dielectric, magnetic) are calculated.\cite{Catlow2010a}

The optimisation of a crystal structure may require hundreds of self-consistent field iterations across a series of ionic configurations.\cite{Payne1992a} 
The most robust approach 
to optimisation is the calculation of an equation of state (EoS) for the material, relating the unit cell dimensions to energy and pressure.\cite{vinet1989universal}
This is based on a series of calculations at differing volumes, where ideally the shape and internal positions are optimised at each point. 
The simplest case is a cubic lattice with high internal symmetry, e.g. rocksalt, where the only degree of freedom is the volume, and computing the EoS reduces to a series of single-point calculations.
For a triclinic cell, the lengths, angles and internal positions in principle all require optimisation.   
While it is sometimes possible to directly optimise the cell volume by \textit{simultaneously} minimising the stress tensor of the unit cell, this approach can run into artifacts when using plane-wave basis sets (i.e. Pulay forces) unless an iterative procedure is employed.\cite{Francis1990}

It has become commonplace to use an `equilibrium' crystal geometry, determined using one exchange-correlation functional within density functional theory,
for a `single-shot' higher-level calculation performed to give a more accurate electronic structure.
 This methodology has been applied to the calculation of properties as diverse as workfunctions, electronic bandgaps, optical properties and defect formation energies.\cite{Walsh2008c,lany2009accurate,shimazaki2009first,Walsh2009g,castelli2012new,Burton2013,Hendon2014a,bhachu2015scalable}
The implicit assumption is that the qualitative behaviour is insensitive to small differences in the local structure.
The approximation will fail where the electronic structure (chemical bonding) of a system
is poorly described at the initial level of theory, e.g. the treatment of Mott insulators such as NiO within
the local-density approximation (LDA).\cite{dudarev2000correlation} 

In this contribution we outline a simple procedure for the rapid volume optimisation (RVO) of crystal structures.
It takes advantage of the similarity in the pressure-volume relationship for a given material between different theoretical approaches, here being exchange-correlation (XC) functionals within density functional theory. 
Where an EoS is known for one functional, the equilibrium volume for another functional can be predicted with reasonable accuracy using a single calculation, and further refined following an iterative procedure.
The approach has particular utility for studies assessing material properties using a range of electronic-structure methods, and for studies employing methods with high computational cost (e.g. hybrid, meta-hybrid and double-hybrid treatments of electron exchange and correlation). 
We validate the approach for four Zn and Pb chalcogenides, and demonstrate its utility in describing the electronic and magnetic structure of one complex semiconductor (\ce{Cu2ZnSnS4}) and one metal-organic framework (HKUST-1), respectively.

\section{Outline of Procedure}
The goal of local crystal-structure optimisation is to minimise all degrees of freedom (cell size, shape and positions) with respect to the total energy of the system. 
It is convenient to employ an EoS based on an energy-volume (E-V) curve, where the remaining degrees of freedom (i.e. shape and positions) are minimised for each volume using standard numerical minimisation approaches (e.g. the conjugate-gradient method).
Kohn-Sham density functional theory (DFT)\cite{Kohn1965b} is one of the most widely used electronic structure techniques for modelling solid-state materials.
Most DFT codes provide optimisation algorithms for this purpose, e.g. the \texttt{ISIF=4} setting in the Vienna Ab initio Simulations Package (VASP)\cite{Kresse2014}, the \texttt{cell\_dofree=`volume'} setting in Quantum-Espresso\cite{baroni2001pwscf} or the \texttt{CVOLOPT} setting in CRYSTAL.\cite{Dovesi2014}

\begin{figure}
\centering
\begin{subfigure}[b]{8cm}
  \includegraphics[width=8cm]{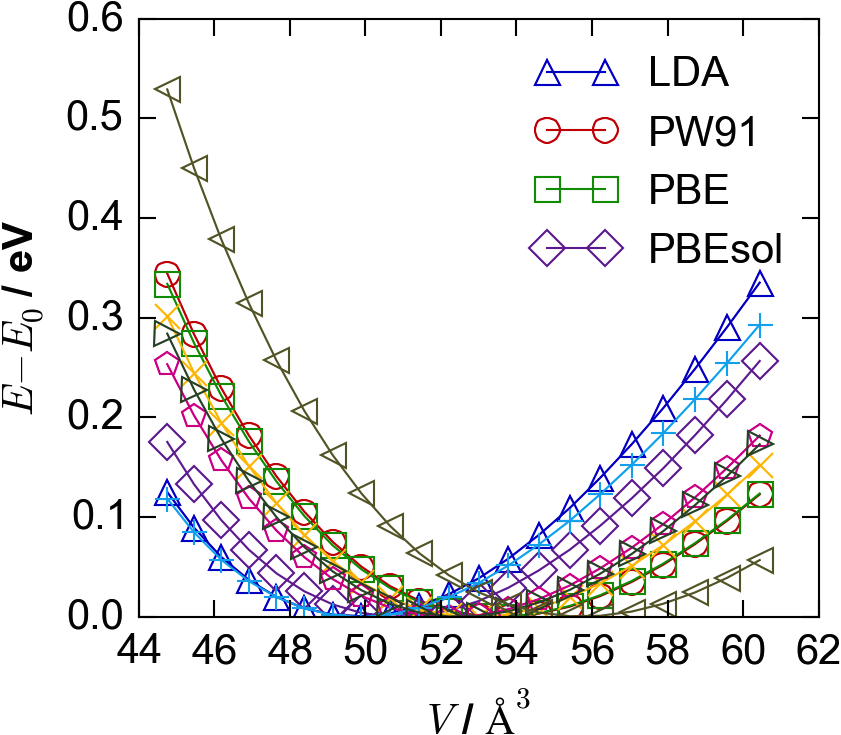}
  \subcaption{
  \label{fig:PbS_EoS_a}}
\end{subfigure}
\begin{subfigure}[b]{8cm}
  \includegraphics[width=8cm]{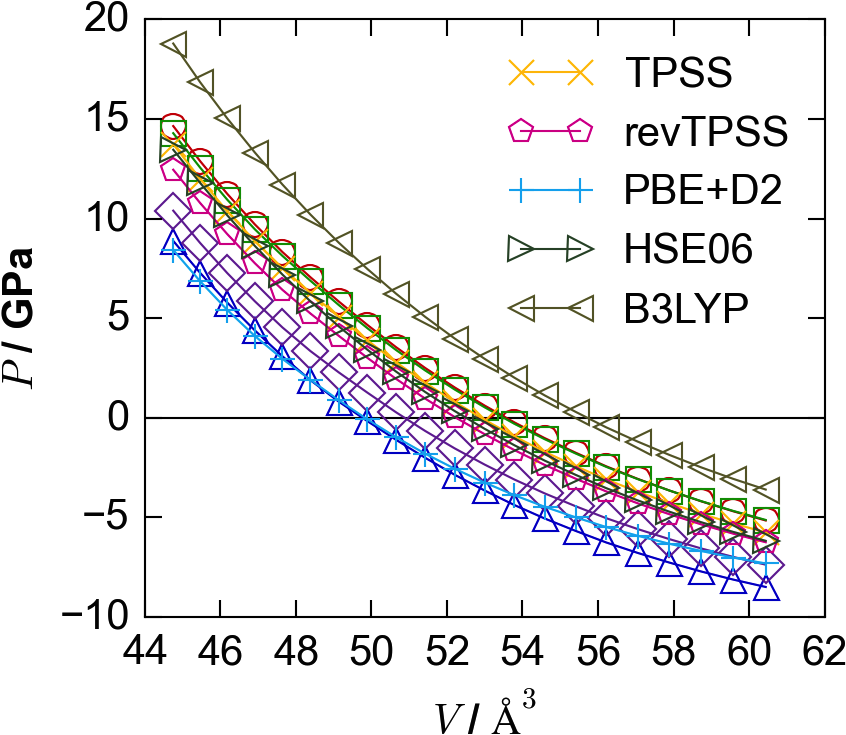}
  \subcaption{
  \label{fig:PbS_EoS_b}}
\end{subfigure}
\caption{Energy-volume and pressure-volume curves computed for PbS using a variety of DFT exchange-correlation functionals.
Markers indicate calculated values, while smooth lines are fits to the Murnaghan equation of state.
\label{fig:PbS_EoS}}
\end{figure}

A superficial resemblance is clear between E-V curves obtained with different exchange-correlation functionals,
with similar shapes but different minima (Figure~\ref{fig:PbS_EoS_a}).
The extent of the similarity becomes apparent when using pressure-volume (P-V) curves (Figure~\ref{fig:PbS_EoS_b}), where ``pressure'' refers to the scalar hydrostatic pressure on the periodic system.
As this pressure $P = - \diff{E}{V}$, the optimal geometries $\diff{E}{V}=0$ are now those intersecting the $P=0$ line.
We note that while these still differ depending on the chosen XC functional, the P-V curves have similar curvature,
with the same approximate slopes about their zero-crossing points.
From these we make our key assumption: 
\emph{the slope of one P-V curve may be used to estimate the crossing point of another}.

For certain beyond-DFT calculation methods, the stress tensor is not computed directly.
However, where the energy is available the hydrostatic pressure may always be estimated with a finite difference:
\begin{equation}
P(V) \approx - \frac{E(V+\delta) - E(V)}{(V+\delta) - V} \label{eqn:finite-difference}
\end{equation}

\begin{figure*}
 \centering
 \includegraphics{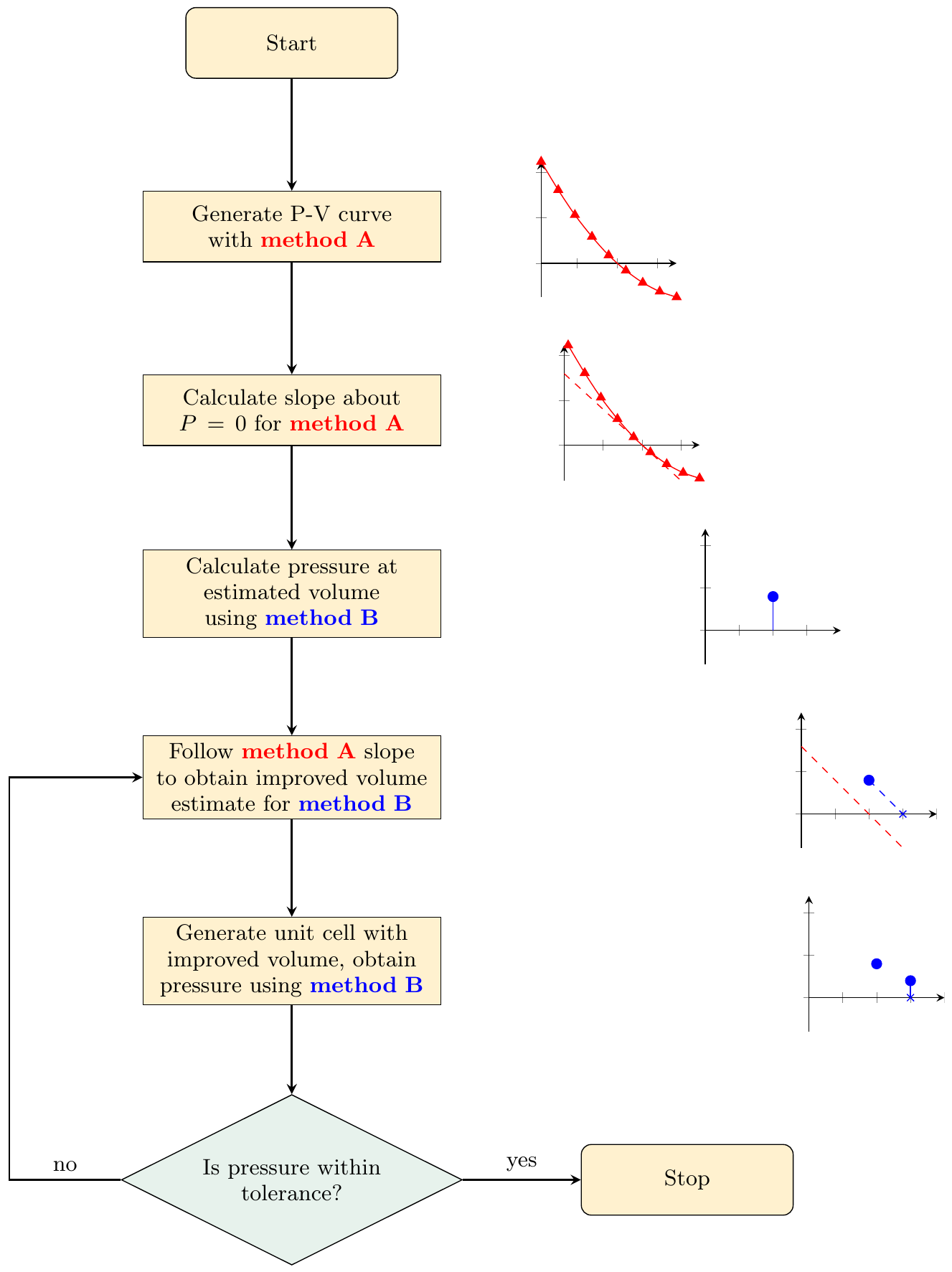}
\caption{Flow chart for the rapid volume optimisation procedure.}
 \label{Image1}
\end{figure*}

The procedure, outlined in Figure~\ref{Image1}, is:

\begin{enumerate}
\item Form a P-V curve using one description of the interatomic interactions (method A).
  This can be achieved  by fitting an EoS to an energy-volume curve.
  If a system-specific set of classical potentials is available, this would be expected to perform very well as they are often fitted to the experimental lattice parameters and elastic properties.
Within DFT, descriptions of electron exchange and correlation within the generalised-gradient approximation (GGA) are suitable,\cite{perdew1992atoms}
    given their low cost and the availability of analytical gradients for the rapid calculation of forces.
    Comparative studies suggest that PBEsol\cite{Perdew2008} gives especially good estimates for the lattice parameters and elastic properties of crystals.\cite{DelaPierre2011,Csonka2009}
\item Calculate the slope about $P=0$ for method A.
  This will form our linear approximation.
  \begin{equation}
    m = \diff{P_A}{V}\Bigr\rvert_{P_A=0}
  \end{equation}
\item Perform a calculation using a second approach (method B), e.g. hybrid DFT with the screened HSE06 functional\cite{Krukau2006},
using an estimated initial volume; this may be the equilibrium volume ($V_0$) for method A.
Convert the resulting stress tensor to a hydrostatic pressure $P_0$. (If no analytical stress tensor is available, use a finite difference as in Eqn.~\ref{eqn:finite-difference}.)
\item Estimate the corrected volume for method B:
  \begin{equation}
    V_1 = V_0 + \frac{P_0}{m}
  \end{equation} \label{step:est}
\item Generate a unit cell with volume $V_1$ (e.g. by interpolating between the previous calculations with method A), and recalculate the desired properties with method B. \label{step:recalc}
\item Iterate steps \ref{step:est}-\ref{step:recalc} until $P$ is acceptably low:
  \begin{equation}
    V_{n+1} = V_n + \frac{P_n}{m}, \quad\quad P_n = f(V_n)
  \end{equation}
\end{enumerate}

\begin{table}
\caption{Equilibrium properties of PbS from the Murnaghan EoS, fitting over a range of functionals: lattice parameter $a$; unit cell volume $V_0$; volume difference $\epsilon_V$ from experimental value; Murnaghan EoS parameters $k_{0}$ and $k^{\prime}_{0}$. $k_0$ is equivalent to Bulk modulus at zero pressure. The experimental lattice constant was obtained from low-temperature neutron powder diffraction data fitted and extrapolated to zero temperature by K. S. Knight (2014).\cite{Knight2014}
\label{tbl:binary_results}}

\begin{ruledtabular}
\begin{tabular}{l d d d d d}
\multicolumn{1}{c}{XC functional} & \multicolumn{1}{c}{ $a$ / \AA} & \multicolumn{1}{c}{$V_{0}$ / \AA$^{3}$} &  \multicolumn{1}{c}{$\epsilon_{V}$ / \% }  & \multicolumn{1}{c}{$k_{0}$}  & \multicolumn{1}{c}{$k^{\prime}_{0}$} \\  
\hline
LDA       &      5.84 & 199.01 &    -3.47  &     65.71 &    4.42  \\
PW91      &      5.99 & 215.21 &     4.38  &     55.26 &    3.98  \\
PBE       &      5.98 & 214.18 &     3.88  &     54.64 &    4.00  \\
PBEsol    &      5.88 & 203.43 &    -1.33  &     61.13 &    4.25  \\
TPSS      &      5.96 & 211.76 &     2.71  &     57.33 &    4.01  \\
revTPSS   &      5.94 & 209.05 &     1.39  &     57.85 &    4.00  \\
PBE+D2    &      5.84 & 199.19 &    -3.39  &     59.93 &    5.02  \\
B3LYP     &      6.06 & 223.02 &     8.17  &     53.20 &    4.07  \\
HSE06     &      5.96 & 210.09 &     1.90  &     59.29 &    4.32  \\
\hline
Experiment\cite{Knight2014}   &     5.91  &  206.17 & \multicolumn{3}{c}{\rule[3pt]{6em}{0.5pt}}  \\
\end{tabular}
\end{ruledtabular}
\end{table}

\section{Error Estimation}
\subsection{Accuracy of linear approximation}
In this approach, a linear fit is used for the pressure-volume relationship:
\begin{align}
P_\text{est} &= aV + b  \\
\frac{\text{d}P_\text{est}}{\text{d}V} &= a. \label{eqn:deriv-a}
\end{align}
This is by no means a conventional equation of state, but may provide a useful approximation when close to the minimum volume.
The standard definition of the bulk modulus,
\begin{equation}
B = -V \frac{\text{d}P}{\text{d}V},
\end{equation}
yields the static bulk modulus $B_0$ when evaluated at the equilibrium volume $V_0$.
\begin{equation}
B_0 = -V_0 \cdot \frac{\text{d}P}{\text{d}V}\Bigr\rvert_{V=V_0}
\end{equation}
As we have assumed this derivative to be constant, we combine with Eqn.~(\ref{eqn:deriv-a}) 
to give a physically meaningful expression of our assumption:
\begin{equation}
\frac{\text{d}P_\text{est}}{\text{d}V} = a = -\frac{B_0}{V_0}. \label{eqn:amazingbulk}
\end{equation}
It is now straightforward to compare this approximate EoS with a more conventional form for solid materials,
estimating an associated error ($\epsilon$).
The simplest case is a system with constant bulk modulus.
\begin{align}
\diff{P}{V} &= -\frac{B_0}{V} \label{eqn:const-b-eos} \\
\epsilon &= P - P_\text{est} \\
\diff{\epsilon}{V} &= \diff{P}{V} - \diff{P_{\text{est}}}{V}
    = \frac{-B_0}{V} - \frac{-B_0}{V_0}  \nonumber \\
    &= B_0 \left( \frac{1}{V_0}-\frac{1}{V} \right) \\
    \epsilon(V) &= \int^V_{V_0} \frac{\text{d}\epsilon}{\text{d}V}\text{d}V
    = B_0 \left[ \frac{V}{V_0} - \ln V \right]^V_{V_0} \\
\epsilon(V) &= B_0 \left( \frac{V}{V_0} -1 +\ln \left( \frac{V_0}{V} \right)  \right) \nonumber \\
    &= B_0 \left( \frac{V-V_0}{V_0} -\ln \left(\frac{V}{V_0}\right) \right)
\intertext{At a typical volume deviation of 5\% (Table~\ref{tbl:binary_results}):}
\epsilon(1.05 V_0) &= B_0 \left( \frac{1.05 V_0-V_0}{V_0} -\ln \left(\frac{1.05 V_0}{V_0}\right) \right) \\
&= B_0 (0.05 - \ln(1.05)) \\
\intertext{where the pressure}
P &= \int^{1.05 V_0}_{V_0} -\frac{B_0}{V} \text{d}V = -B_0 \ln \left( \frac{1.05 V_0}{V_0} \right)
\intertext{and hence the fractional error $\frac{\epsilon}{P}$ is -2.5\%. \nonumber }
\end{align}

\begin{figure}



\definecolor{jctc-green}{RGB}{15,122,68}
\definecolor{jctc-orange}{RGB}{255,241,207}

\begin{tikzpicture}
  \begin{axis}[
      xlabel={$\frac{V}{V_0}$},
      ylabel={$\epsilon$ / $B_0$},
      xmin=0.8,
      xmax=1.2,
      ymin=0,
       x tick label style={
         /pgf/number format/.cd,
            fixed,
            fixed zerofill,
            precision=1,
        /tikz/.cd
    },
           y tick label style={
         /pgf/number format/.cd,
            fixed,
            fixed zerofill,
            precision=2,
        /tikz/.cd
    }
    ]
    \addplot[jctc-green,domain=0.8:1.2]
    {x-1+ln(1/x)};
    \addlegendentry{$\frac{\textrm{d}P}{\textrm{d}V}=-\frac{B_0}{V}$};
    \addplot[red, domain=0.8:1.2]
    {(x-1) + (1/5)*(x^(-5)-1))};
    \addlegendentry{Murnaghan ($k^\prime_0 = 5$)};
    \end{axis}
\end{tikzpicture}

\caption{Divergence of the linear approximation from more complex equations of state. The error $\epsilon = P_\mathrm{est} - P_\mathrm{EOS}$ is given in units of the static bulk modulus.  
\label{fig:error}}
\end{figure}
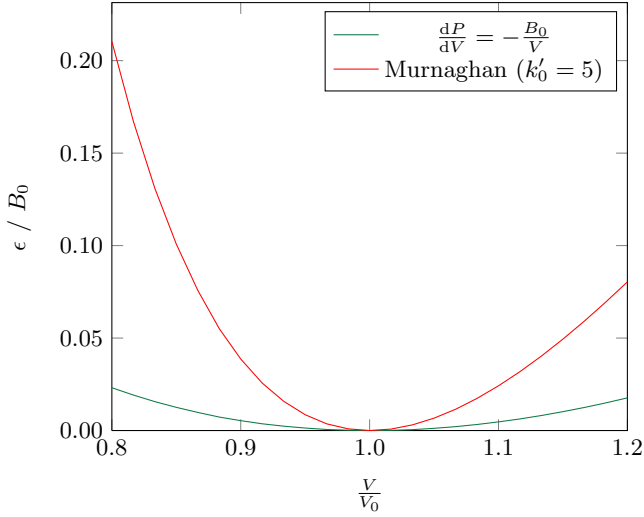

Moving to an improved, while still relatively simple, EoS, the Murnaghan EoS adds a parameter, effectively giving a linear volume dependence to $B_0$.\cite{Murnaghan1944}
Taking its derivative form (Eqn. \ref{eqn:murnaghan-prime}), we improve our error estimate:
\begin{align}
\diff{P}{V} &= - \frac{k_0}{V}\left(\frac{V_0}{V}\right)^{k^\prime_0} \label{eqn:murnaghan-prime} \\
\diff{\epsilon}{V} &= \diff{P}{V} - \diff{P_\text{est}}{V} = 
    \frac{k_0}{V}\left(\frac{V_0}{V}\right)^{k^\prime_0} - \frac{-B_0}{V_0} \\
\intertext{We can relate $B_0$ to the Murnaghan parameters $k_0, k^{\prime}_0$ by finding the slope at $V_0$:}
B_0 &= -V_0 \cdot \diff{P}{V}\Bigr\rvert_{V=V_0} = \frac{V_0}{V_0}k_0\left(\frac{V_0}{V_0}\right)^{k^\prime_0} = k_0 \\
\diff{\epsilon}{V} &= k_0 \left(\frac{1}{V_0} - \frac{V_0^{k^\prime_0}}{V^{k^\prime_0 + 1}} \right) \\
 \epsilon(V) &= \int^V_{V_0} k_0 \left( \frac{1}{V_0} - \frac{V_0^{k^\prime_0}}{V^{k^\prime_0+1}} \right) \text{d}V \\
&= k_0 \left[ \frac{V}{V_0} + \frac{1}{k^\prime_0} \left(\frac{V_0}{V} \right)^{k^\prime_0} \right]^V_{V_0}\\
\end{align}
\begin{align}
 \epsilon(V) &= k_0 \left( \frac{V}{V_0} + \frac{1}{k^\prime_0} \left( \frac{V}{V_0}\right)^{-k^\prime_0} -1 - \frac{1}{k^\prime_0}\right) \\
&= k_0 \left(\frac{V}{V_0}-1\right) + \frac{k_0}{k^\prime_0} \left( \left( \frac{V}{V_0}\right)^{-k^\prime_0} -1 \right)
\end{align}
Plotting these error estimates in Figure~\ref{fig:error}, we find that the error in pressure is less than 10\% of the static bulk modulus for volumes ~10\% from the optimum, 
given a material that obeys the Murnaghan EoS with a typical value $k^{\prime}_0 = 5$. 
For smaller values of $k^{\prime}_0$ (i.e. closer to the fixed-bulk-modulus model), the errors are greatly reduced.
In any case, the linear approximation appears to be sufficiently accurate for a stable optimisation process.

\subsection{Dependence on accuracy of EoS}
Returning to the simplistic EoS of Eqn.~\ref{eqn:const-b-eos}
\begin{align}
P &= - \int^V_{V_0} \frac{B_0}{V} \textrm{d}V \\
&= B_0 \ln{\left( \frac{V_0}{V} \right)}, \\
\intertext{we examine the residual pressure $P_1$ following a single step of RVO from an initial volume $V_i$,}
V_1 &= V_i + \frac{P_B (V_i)}{m_A} = V_i - P_B(V_i)\frac{V_{0,A}}{B_{0,A}} \\
&= V_i - B_{0,B} \ln \left( \frac{V_{0,B}}{V_i} \right) \cdot \frac{V_{0,A}}{B_{0,A}} \\
P_1 &= B_{0,B} \ln \left( \frac{V_{0,B}}{V_1}\right) \\
\frac{P_1}{B_{0,B}}&= - \ln \left[ \frac{V_i}{V_{0,B}} - \frac{B_{0,B} V_{0,A}}{B_{0,A}V_{0,B}} \ln \left(\frac{V_{0,B}}{V_i} \right) \right]. \\
\intertext{We note that $\ln(x) \approx x-1$ for $x$ close to 1,
and hence the residual pressure is approximately linear with respect to the error in initial volume estimate.
The term $\frac{B_{0,B} V_{0,A}}{B_{0,A}V_{0,B}}$ indicates a smaller linear dependence on the similarity of the EoS for method A and method B.
Moving to the Murnaghan EoS,}
P_1 &= \frac{k_{0,B}}{k^{\prime}_{0,B}}\left(\left( \frac{V_{0,B}}{V_1} \right)^{k^\prime_{0,B}} -1 \right) \\
&= \frac{k_{0,B}}{k^\prime_{0,B}} \left[ \left( \frac{V_{0,B}}{V_i + P_B(V_i)\frac{V_{0,A}}{-k_{0,A}}} \right)^{k^\prime_{0,B}} -1 \right] \\
\frac{P_1}{\frac{k_{0,B}}{k_{0,B}^\prime}}&=  \left( \frac{V_{0,B}}{V_i - \frac{V_{0,A}}{k_{0,A}} \cdot \frac{k_{0,B}}{k^\prime_{0,B}} \left(\left( \frac{V_{0,B}}{V_i}\right)^{k^\prime_{0,B}}  -1 \right)} \right)^{k^\prime_{0,B}} -1 \\
\frac{P_1}{\frac{k_{0,B}}{k_{0,B}^\prime}}&= \left( \frac{V_i}{V_{0,B}} - \frac{V_{0,A} k_{0,B}}{V_{0,B} k_{0,A}}\cdot \frac{1}{k^\prime_{B,0}} \left( \left( \frac{V_{0,B}}{V_i} \right)^{k^\prime_{0,B}} -1 \right) \right)^{-k^\prime_{0,B}} - 1 
\end{align}
Again, the pressure is dominated by the initial position, with a smaller contribution from the difference between EoS ``stiffness'' and volume minima.
The non-linearity of this relationship follows the non-linearity of the true EoS through the exponent $k^\prime_{0,B}$.
We conclude therefore that the performance of the first step is equally sensitive to percentage differences in equilibrium volume and bulk modulus between method A and method B.
Convergence is impacted by the non-linearity of the EoS, but not by how accurately this non-linearity is reproduced by method A.

\section{Methods}
\subsection{Electronic Structure Calculations}
Studies have been carried out on the binary chalcogenides PbS, PbTe, ZnS, and ZnTe as well as the quaternary semiconductor \CZTS{} and an organic-inorganic hybrid material HKUST-1.

All DFT calculations on the binary chalcogenides were carried out with VASP\cite{PhysRevB.47.558}
using the two-atom primitive face-centred cubic unit cells.
We employed projector-augmented wave (PAW) frozen-core potentials\cite{PhysRevB.50.17953,PhysRevB.59.1758} treating the outermost s and p electrons of S, Te, and Pb and the outermost s, p, and d electrons of Zn explicitly as valence. 
For consistency, we used the LDA PAW potential set.
The PAW potentials are highly transferable and tests showed a very weak dependence of the resulting optimised electronic and crystal structures.  
A plane-wave kinetic-energy cutoff of 550~eV was employed in all these calculations, and the Brillouin zone was sampled with an $8\times8\times8$ $\Gamma$-centered Monkhorst-Pack mesh.\cite{PhysRevB.13.5188}
During electronic minimisation, the wavefunctions were optimised to an energy tolerance of $10^{-8}$~eV. 
These parameters were found to be sufficient to converge the absolute total energies to within 1~meV~atom$^{-1}$, and the stress tensors to well within 1~kbar (0.1 GPa).

The simplicity of the binary systems allowed us to test a wide range of functionals, spanning different ``levels'' of approximations to the exchange-correlation potential.\cite{Perdew2001} 
As a baseline, we took the local-density approximation (LDA) with the Perdew-Zunger parameterisation of the correlation energy.\cite{PhysRevB.23.5048} 
Calculations using the generalised-gradient (GGA) approximation were performed with the Perdew-Wang 91 (PW91)\cite{PhysRevB.44.13298,PhysRevB.46.6671,PhysRevB.48.4978.2} and Perdew-Burke-Enzerhof (PBE)\cite{PhysRevLett.77.3865} functionals, plus the variant of PBE revised for solids (PBEsol).\cite{Perdew2008} 
To complement this set of functionals, we also tested the Grimme D2 dispersion correction to PBE.\cite{JCC:JCC20495} 
``Meta-GGA'' calculations were carried out using the Tao-Perdew-Staroverov-Scuseria (TPSS) functional\cite{PhysRevLett.91.146401} and the subsequent revision of Perdew et al. (revTPSS).\cite{PhysRevLett.103.026403} 
Finally, we tested two hybrids, \emph{viz.} the popular HSE06\cite{Krukau2006} and B3LYP\cite{Becke1993} functionals. 
For each material and functional, we calculated an E-V curve by adjusting the lattice parameter to yield 21 volumes about the experimental 300 K lattice parameters reported in Refs. \nocite{ADFM:ADFM201300722}\citenum{ADFM:ADFM201300722} and \nocite{McIntyre:a18377}\citenum{McIntyre:a18377} covering a range of approx. $\pm 5 \%$.  
We note that, as a result of the high symmetry of these systems, the lattice parameter is the only degree of freedom, and thus relaxation of the cell shape and internal positions was not required.

For \CZTS{} (Section~\ref{sec:czts}), energy-volume curves were formed from all-electron DFT calculations using the FHI-aims code.\cite{Blum2009,Havu2009}
These calculations employed numerically-tabulated atom-centered basis functions (the `tight' defaults were used, which correspond to expected convergence of $< 10$ meV per atom) and evenly-spaced \kpoint{} grids.
Additional hybrid (HSE06) DFT calculations and primitive-cell optimisations used VASP with the PAW-PBE potential set and a 500~eV cutoff for the plane wave basis set.
All calculations on \CZTS{} sampled the Brillouin zone with $6\times6\times6$ $\Gamma$-centered \kpoint{} grids.

For the Cu-based metal-organic framework HKUST-1, calculations were again performed with the VASP code, considering only the point $\Gamma$ in reciprocal space due to the large size of unit cell.
Owing to the presence of the open-shell Cu(II) ion (d$^9$ configuration) all calculations were spin-polarised, and a range of magnetic structures were tested as discussed in Section~\ref{sec:results-hkust}.
The PBEsol and HSEsol functionals were used along with the PAW-PBE potential set.
Here `HSEsol' refers to a modification of HSE06, with PBEsol replacing PBE as the local exchange-correlation functional.\cite{Schimka2011}
Due to the complexity of the crystal structure, only three energy-volume points were included in the EoS and a single iteration of RVO
was performed to recover the ground-state HSEsol structure.
A slightly different procedure was followed in this case: a quadratic E-V curve was fitted to the three PBEsol points.
The initial HSEsol calculation was carried out at the fully-optimised PBEsol point, and the E-V curve was followed assuming a constant pressure offset to estimate the equilibrium volume for HSEsol.
(This application was the first chronologically, and led to the development of the Murnaghan fitting procedure.)

\subsection{Implementation}
The RVO method was implemented and tested with code written in Python 2.7.3, using the standard library and Numpy/Scipy/Matplotlib.\cite{Oliphant2007,Scipy2001,Hunter2007}
(The code is freely available; details in Supporting Information.)
Non-linear fitting to the Murnaghan EoS uses the {\tt curve\_fit } routine in the Scipy Python library, which accesses Minpack, an open-source Fortran library.\cite{Scipy2001}
This implements least-squares fitting with the Levenberg-Marquardt algorithm.\cite{Marquardt1963}
Initial guesses of 50.0 eV $\text{\AA}^{-3}$ and 5.0 were used for the $k^\prime$ and $k_0^\prime$ parameters, respectively.

\section{Results}

\subsection{II-VI Binary Chalcogenide Semiconductors}
\subsubsection{Simulated application across a range of methods}
\begin{figure}
  \begin{center}
    \includegraphics{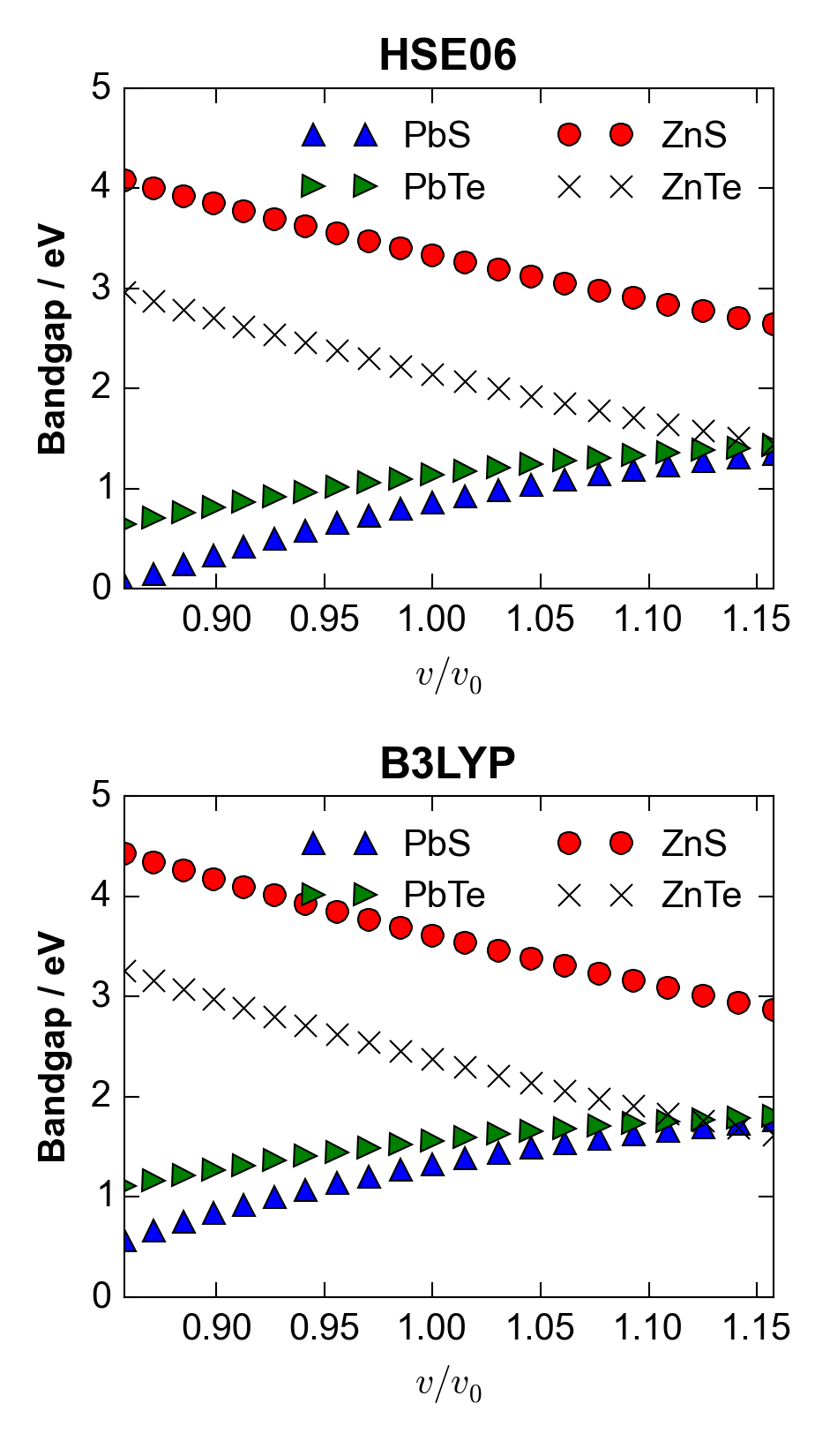}
  \end{center}
\caption{Volume-dependence of calculated direct bandgaps at $\Gamma$ (ZnS, ZnTe) and $\mathrm{X}$ (PbS, PbTe) with the HSE06 and B3LYP hybrid DFT functionals. 
Results as a function of volume (temperature) for PbTe have previously been reported in Ref.~\citenum{Skelton2014}.
The behaviour is characteristic of a positive and negative bandgap deformation potential for the Pb and Zn semiconductors, respectively.
The volume expansion range of 0.86--1.16 corresponds to lattice parameter expansions and contractions of 5\% in each dimension. Markers indicate the calculated values.
\label{fig:v-bandgaps}}
\end{figure}

For PbS there is a significant range in equilibrium lattice parameters for different exchange-correlation functionals,
corresponding to a maximum volume difference of over 10\%, between the LDA and B3LYP calculations (Figure~\ref{fig:PbS_EoS}).
Values are tabulated in Table~\ref{tbl:binary_results}, and compared to a recent low-temperature study by K. S. Knight.\cite{Knight2014}
The computed values are similar, but slightly different, to the computational results reported by Hummer et al.\cite{Hummer2007}
Direct bandgaps were also calculated at each volume expansion, at the special \kpoint{} $\mathrm{X}$ for the lead compounds and at $\Gamma$ for the zinc compounds.
(Note that PbS and PbTe have another, smaller direct bandgap at the $L$ point.)
It can be seen in Figure~\ref{fig:v-bandgaps} that over the lattice-parameter expansion and contraction of up to 5\%,
the bandgaps vary by around 1 eV, with the direction of change depending on the structure type and chemistry.
In this case, using an LDA-predicted geometry for a `single-shot' B3LYP calculation would lead to a difference in bandgap of $\sim 0.4$~eV compared to that at the equilibrium geometry for B3LYP.

\begin{figure*}
  \centering
  \includegraphics[width=16cm]{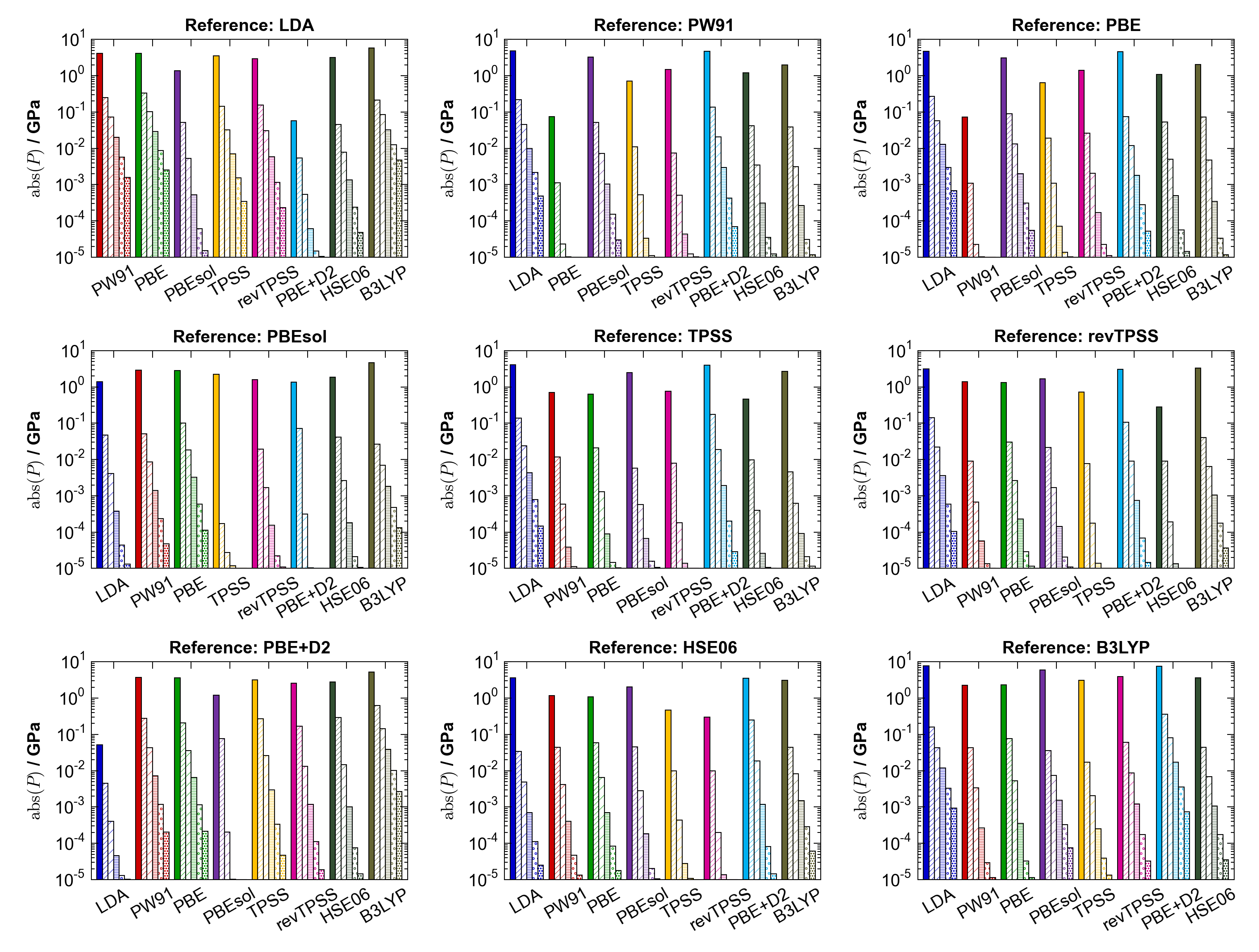}
  \caption{Residual pressures following six iterations of volume optimisation of PbS using different ``reference''/''test'' combinations of exchange-correlation functionals.
  The ``reference'' refers to the functional used for the EoS (``method A''), while the groups of bars correspond to the functionals used in simulated single-point calculations (``method B'').
  Note that the pressure is presented on a logarithmic scale.
    \label{fig:convergence-functionals}}
\end{figure*}

An iterative application of the RVO procedure was then simulated from the data.
The Murnaghan EoS (Eqn.~\ref{eqn:Murnaghan}) in its integrated form (Eqn.~\ref{eqn:MurnaghanE}) was fitted to each E-V curve from DFT calculations.
This allowed energy and pressure to be calculated for arbitrary volumes without carrying out additional DFT calculations.
\begin{align}
  P &= \frac{k_0}{k^\prime_0} \left( \left(\frac{V_0}{V}\right) ^{k^\prime_0} -1 \right) \label{eqn:Murnaghan} \\
  E &= E_0 + k_0  V_0  \biggl( \left(\frac{V}{V_0}\right) ^{1 - k^\prime_0} \cdot \frac{1}{k^\prime_0 (k^\prime_0 - 1)} \notag \\
            & \hspace{4cm}  + \frac{V}{k^\prime_0 V_0} - \frac{1}{k^\prime_0 - 1}  \biggr) \label{eqn:MurnaghanE}
\end{align}
The quality of these fits was sufficient for this exercise, with RMS fitting
errors of $<1$~meV.  
Fitting parameters and full data are included in ESI.  
For each ``test functional''-``reference functional'' pair, the minimum volume
(corresponding to the fitting parameter $V_0$) of the reference functional was
taken as the initial volume guess, and an external pressure calculation
modelled by evaluating the pressure at this volume using the EoS for the test
functional.
This refined pressure and volume was then used as the basis for further iterations.
The external pressure over successive iterations is shown for PbS in
Figure~\ref{fig:convergence-functionals} for each combination of functionals;
convergence is rapid with the residual pressure dropping almost logarithmically with subsequent steps,
typically by a factor of $\sim 10^3$ in three steps.

\subsubsection{Comparison with a standard optimisation procedure}
In general terms, a direct optimisation with method B will take $N_{\mathrm{opt},B}$ steps, each requiring an average computing time $t_B$, to converge to the equilibrium volume.
Constructing an EoS for RVO using method A requires $N_{\mathrm{EoS},A}$ optimisations, which, as for method B, take $N_{\mathrm{opt},A}$ steps of time $t_A$.
We note that in most cases $N_{\mathrm{opt},B}$ will be larger than $N_{\mathrm{opt},A}$, since the direct optimisation with method B must adjust the internal coordinates, cell shape and volume, while the optimisation with method A needs only to optimise the internal coordinates and the shape.
Subsequent application of $N_{\mathrm{RVO}}$ iterations of the algorithm then requires $1 + N_{\mathrm{RVO}}$ single-point calculations using method B, each again requiring $t_B$ time.
RVO is expected to be more efficient than a direct optimisation with method B if the following inequality holds:
\begin{equation}
N_{\mathrm{EoS},A} N_{\mathrm{opt},A} t_A + \left(1 + N_\mathrm{RVO} \right) t_B < N_{\mathrm{opt},B} t_B .
\end{equation}

The cubic systems considered in this section, for which the cell volume is the only degree of freedom, represent a special case where $N_{\mathrm{opt},A}=1$.
We assume that energy gradients are available with method B, and that the optimisation algorithm would converge in three steps, i.e. $N_{\mathrm{opt},B}=3$.
This is reasonable if a good estimate of the starting volume is available, such as a room-temperature lattice constant.
The inequality simplifies to
\begin{equation}
N_{\mathrm{EoS},A} t_A + \left(1 + N_{\mathrm{RVO}} \right) t_{B} < 3 t_{B};
\end{equation}
it can be seen that RVO will outperform a direct optimisation if a suitable pressure is obtained after one iteration while $t_A< \frac{t_B}{ N_{\mathrm{EoS},A}}$.

As a concrete example, we compared a direct optimisation of PbS with HSE06 to an optimisation with RVO using PBEsol and HSE06 as method A and method B, respectively.
The initial structure for both optimisations was the experimentally-measured room temperature volume, and an eleven-point EoS for RVO was computed about this value using PBEsol. 
The direct optimisation used a quasi-Newton algorithm as implemented by VASP (with the input tag ``IBRION=1'').
Both sets of calculations were performed on a dual-CPU Intel Xeon workstation with 12 physical cores and 64 Gb RAM, allowing the timings to be compared directly. 
The comparison is given in Table~\ref{tbl:shootout}.

\begin{table}
  \caption{
    Comparison of a direct HSE06 volume optimisation and one and two iterations of the RVO algorithm in determining the equilibrium volume of PbS.
    For each step, the total time for each step is recorded alongside the cell volume and pressure after the cycle where appropriate. 
For the direct optimisation, the timings of the three steps are printed alongside the total for the complete calculation, so the latter includes additional operations such as setup time and is slightly longer than the sum of the three electronic minimisations.
\label{tbl:shootout}}
  \begin{ruledtabular}
  \begin{tabular}{l l d d d}
    \multicolumn{1}{l}{\text{Algorithm}} & \multicolumn{1}{l}{\text{Step}} &
    \multicolumn{1}{l}{\text{$t$ / s}} & \multicolumn{1}{l}{\text{$V$ / $\AA^3$}} & \multicolumn{1}{l}{\text{$p$ / kbar}} \\
 \hline
    \multirow{4}{*}{Direct (HSE06)} & 1 & 6669 & 52.21 &  3.42 \\
                                    & 2 & 5808 & 52.63 & -1.45 \\
                                    & 3 & 4509 & 52.50 & -0.10 \\
    \cline{2-5}                  & Total & 17038  & &       \\
    \hline
    \multirow{7}{*}{RVO}    & PBEsol EoS & 47  & &             \\
                            & HSE06 1    & 8234 & 52.21 & 3.39 \\
                            & HSE06 2    & 6754 & 52.49 & 0.15 \\
   \cline{2-5}           & Total (1 iter) & 15035 & &       \\
                            & HSE06 3    & 6701 & 52.50 & 0.03 \\
   \cline{2-5}
                            & Total (2 iters) & 21736 & &      \\
  \end{tabular}
  \end{ruledtabular}
\end{table}

In this test, a single-point calculation with HSE06 was on average 150 times more expensive than a calculation with PBEsol;
this is both due to the higher complexity of non-local hybrid functionals compared to semi-local GGA methods, 
and to the different scaling properties with the number of \kpoints{} used to sample the Brillouin zone.
With the force convergence criterion of $10^{-2}$ eV $\AA^{-1}$ for direct optimisation, the pressure was reduced to -0.1 kbar from an initial pressure of 3.42 kbar in three steps, taking 4.75 hrs on our test hardware. A single iteration of RVO yielded a pressure of 0.15 kbar in 4.18 hrs, while a second iteration yielded 0.03 kbar in a total time of 6 hrs.

It can be seen from the data in Table~\ref{tbl:shootout} that the direct optimisation takes on average less time per force calculation than RVO; the procedure implemented in VASP re-uses calculated wavefunctions to speed up the convergence of the second and third steps.
In this case, we found that one of the conjugate-gradient electronic-minimisation cycles during the first single-point calculation for the RVO algorithm took some 1500s longer than both the other steps in this series and the first step of the quasi-Newton volume optimisation,
despite the latter being notionally an identical calculation.
This artefact contributes significantly to the cost of the single-iteration RVO calculation.

Nonetheless, even for this relatively simple test case, useful savings in computing time could potentially be obtained in practice with RVO.
Given the poor scaling of computational cost with system size when using advanced electronic-structure methods, we would expect more substantial savings for more complex unit cells. 
This would also be true in systems where direct optimisation requires the minimisation of a larger number of degrees of freedom, leading a larger number of steps with method B, which 
is the subject of the following case studies.

\subsection{Quaternary Sulfide \ce{Cu2ZnSnS4}}
\label{sec:czts}
\CZTS{} (CZTS) is an attractive light-absorbing material for thin-film photovoltaics, with a direct bandgap and consisting of earth-abundant components,
which has attracted significant experimental\cite{Schorr2011b,Siebentritt2012a,Scragg2012d,Gunawan2014,Scragg2014} and computational\cite{Chen2009,Botti2011,Persson2010,Walsh2012a,Skelton2015} research effort.
In the search for new materials for solar energy conversion, the prediction of accurate bandgaps from first-principles is a serious challenge and CZTS represents a suitable case for probing the effect of crystal structure.

An initial structure for CZTS in the kesterite phase, optimised with PBEsol, was obtained during previous work.\cite{Jackson2014}
This was reduced from a conventional 16-atom body-centered-tetragonal cell with $\bar{\textrm{I}}4$ symmetry to the corresponding 8-atom primitive cell using Spglib.\cite{Spglib2009}
A set of seven structures was obtained for both cells by multiplying each lattice vector by a scale factor from 0.97--1.03 and performing a local optimisation of the atomic positions, this forming the ``method A'' energy-volume curves.
In addition to this isotropic scaling, an additional set of structures were calculated including optimisation of the cell shape (i.e. the tetragonal c/a ratio) for each volume point.
The iterative RVO procedure was followed in order to minimise the pressure and obtain a more accurate electronic structure using the HSE06 functional.
The results are given in Table~\ref{tbl:CZTS_results}; pressure minimisation was rapid in all cases, decreasing logarithmically with each step.

We note that the resulting lattice parameters from these calculations, especially those using the primitive cell, are very close to both the experimental lattice parameter \textit{a}=5.427~\AA{} and theoretical \textit{a}=5.448~\AA{} reported by Paier et al. following a conventional optimisation procedure with a variant of the HSE functional.\cite{Paier2009} We also note a bandgap shift of almost 0.1 eV when the E-V curve was provided by a non-isotropic set of primitive lattices.

\begin{table*}
  \begin{center}
    \caption{Results from application of RVO to \CZTS{}, using the HSE06 functional and a PBEsol-derived E-V curve.
      The unit of pressure $P$ is kbar ($10^8$ Pa) and volumes $V$ are given in \AA$^3$.
      $a$ is the lattice parameter in \AA{};
      these are calculated as a mean over the $a$ and $b$ vectors (crystal symmetry is not enforced in FHI-AIMS).
      $E_g$ is the electronic bandgap in eV taken from the Kohn-Sham eigenvalues at the $\Gamma$-point.
      The methods in parentheses refer to the process by which the E-V curve was generated;
      isotropic expansion energies with FHI-aims and volume-conserving relaxations with VASP.
      Iteration ``2*'' is the data from a final set of calculations, where the internal atomic positions are relaxed while maintaining the unit cell from iteration 2.
      \label{tbl:CZTS_results}}
    \begin{ruledtabular}
      \begin{tabular}{ddddddddddddd}
 & \multicolumn{4}{c}{Conventional cell}  & \multicolumn{4}{c}{Primitive cell}  & \multicolumn{4}{c}{Primitive cell} \\
\multicolumn{1}{c}{\text{Iteration}} & \multicolumn{4}{c}{(Isotropic expansion)}  & \multicolumn{4}{c}{(Isotropic expansion)}  & \multicolumn{4}{c}{(Constrained relaxation)} \\
 & \multicolumn{1}{c}{$P$} & \multicolumn{1}{c}{$V$} & \multicolumn{1}{c}{$a$} & \multicolumn{1}{c}{$E_g$} & \multicolumn{1}{c}{$P$} & \multicolumn{1}{c}{$V$} & \multicolumn{1}{c}{$a$} & \multicolumn{1}{c}{$E_g$} & \multicolumn{1}{c}{$P$} & \multicolumn{1}{c}{$V$} & \multicolumn{1}{c}{$a$} & \multicolumn{1}{c}{$E_g$}   \\
\hline
0 & 22.82 & 309.12 &  5.38 & 1.26 & 17.49 & 155.43 &  5.38 & 1.23 &  17.49 & 155.43 & 5.38 & 1.23 \\
1 & -1.23 & 318.03 &  5.40 & 1.18 & -0.64 & 158.87 &  5.42 & 1.15 &  -1.35 & 159.00 & 5.44 & 1.13 \\
2 & 0.00 & 317.56  &  5.40 & 1.19 & -0.01 & 158.74 &  5.42 & 1.15 &   0.02 & 158.73 & 5.43 & 1.14 \\
\hline
2* & 7.46 & 317.56 & 5.40 & 1.49 & 10.17 & 158.74  & 5.42 & 1.48  & 10.31  & 158.73 & 5.43 & 1.47 \\
      \end{tabular}
    \end{ruledtabular}
\end{center}
\end{table*}

This case also highlights the importance of internal structure optimisation.
After two steps of optimisation using the E-V curve further calculations were carried out, employing the HSE06 exchange-correlation functional, where the internal atomic positions were relaxed while fixing the unit cell. As shown in the table, these lead to an increase in the absolute pressure, but also a considerable improvement in the bandgap estimation compared to experimental measurements.
Previous electronic structure studies have show that the bandgap of CZTS is highly sensitive to the S positions, which correspond to deviations  away from the ideal tetrahedral coordination environment.\cite{Botti2011}
In the ideal kesterite crystal structure, the metal nuclei all occupy high symmetry Wyckoff positions (2\textit{a} and 2\textit{c} by Cu, 2\textit{d} by Zn, and 2\textit{b} by Sn). 
However, the sulfur anions occupy the lower symmetry 8\textit{g} positions, with \textit{x}, \textit{y} and \textit{z} displacement parameters.
The change of $\sim 0.3$~eV in the bandgap following further optimisation (Table~\ref{tbl:CZTS_results}) emphasises the importance of internal relaxations for quantitative studies of electronic structure.

\subsection{Metal-organic Framework HKUST-1}
\label{sec:results-hkust}
In 1999, Williams and co-workers isolated Cu$_{3}$(btc)$_2$ (HKUST-1).\cite{Chui1999}
Since then this material has been widely studied in the field metal-organic frameworks (MOFs), with possible applications in catalysis, ionic and electrical conductivity, photovoltaics, batteries and gas capture.\cite{Hendon2015a}
First-principles calculations of MOF properties have traditionally posed challenges for computational chemists because they combine large unit cells with complex organic and inorganic components.

HKUST-1 features an additional layer of complexity: it is composed of Cu-Cu ``paddlewheel'' inorganic regions, where each Cu(II) atom is associated with an unpaired electron and each paddlewheel is antiferromagnetically (AFM) coupled in the ground state configuration.\cite{Zhang2000,Poppl2008}
 The magnetic interactions are highly sensitive to the Cu-Cu separation.
 Moreover, previous studies\cite{Butler2014a,Butler2014c}
have shown that the ionisation potentials and bandgaps of porous materials are sensitive to cell pressure and volume, 
 similar to some of their inorganic counterparts.
HKUST-1 represents an extreme case, where deviations from the equilibrium Cu-Cu distance result in spin flipping and formation of a ferromagnetic (FM) state, which impacts the electronic structure.

\begin{figure*}
 \includegraphics[scale=0.35]{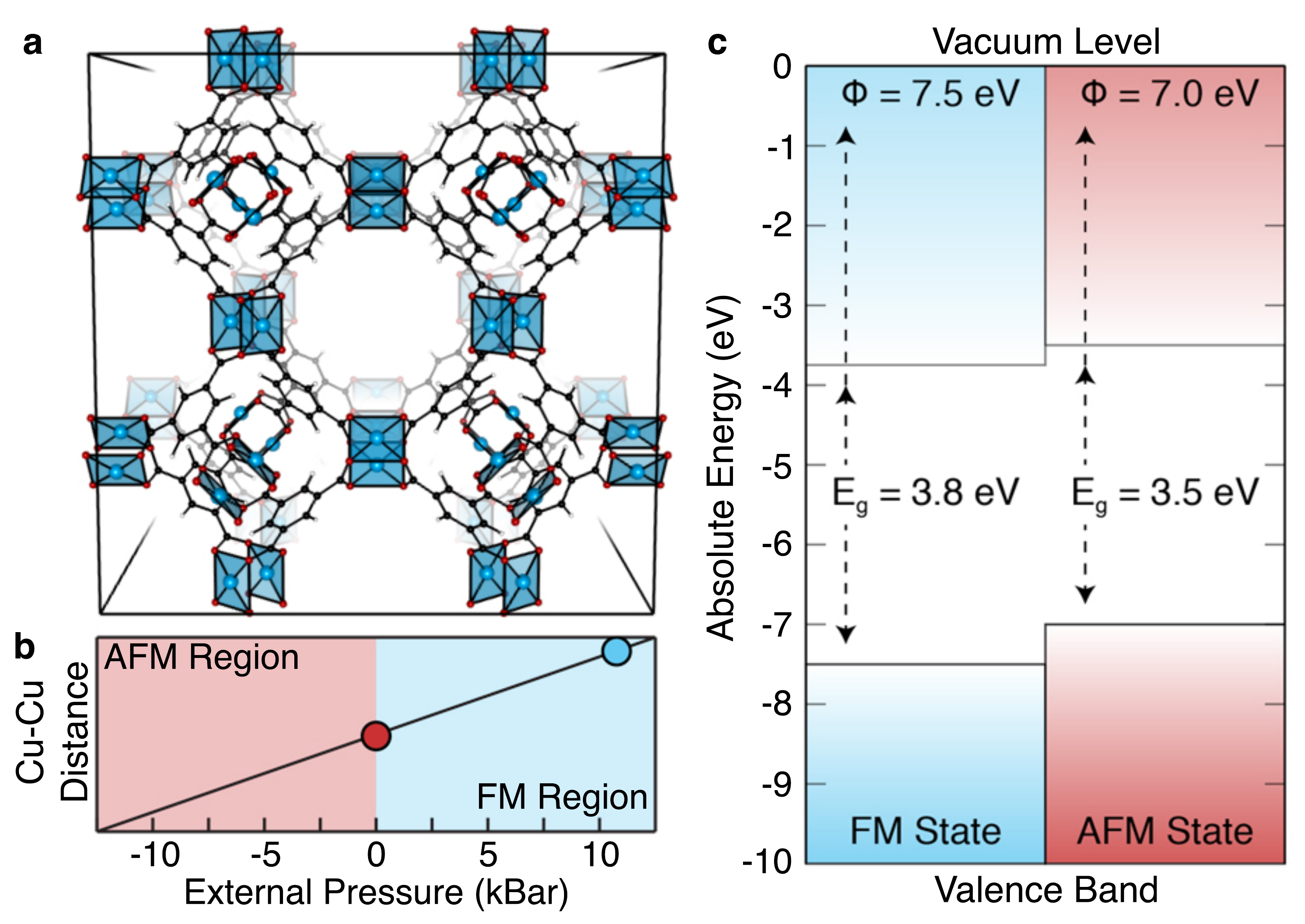}
 \caption{
  (a) HKUST-1 features a periodic array of 32 Cu(II)-Cu(II) paddlewheels per crystallographic unit cell.
  (b) The favoured magnetic structure depends on the Cu-Cu separation:
     antiferromagnetic (AFM) and ferromagnetic (FM) states are accessible. 
  (c) The valence band energy (ionisation potential) is sensitive to the magnetic structure 
  	  (calculated using the procedure outlined in Ref.~\citenum{Butler2014a}).
    A `single-shot' HSE06 calculation on the PBEsol structure favours the FM state (blue), 
    whilst the corrected structure favours the experimentally observed AFM state (red).}
    \label{fig:hkust-results}
\end{figure*}

 Typically, PBEsol-optimised structures agree with low-temperature experimental measurements of MOFs to within 1 \%.
 This is the case here, and PBEsol also reproduces the correct AFM state.
However, a single-point HSE06 calculation on the PBEsol structure yields an incorrect FM ground-state,
as shown in Fig.~\ref{fig:hkust-results}, with an associated HSE06 cell pressure of -13.98 kbar (Table~\ref{tbl:hkust}).
 In order to recover an accurate HSE06 crystal structure (and the associated correct magnetic structure), 
 a single iteration of RVO was required.
 Notably, there is not only a magnetic difference, but also a significant difference in predicted electronic bandgap 
 and workfunction.

 \begin{table}
   \begin{center}
     \caption{
       Results from volume optimisation of HKUST-1. Residual pressure $P$ at each step, energies of valence band maximum (VBM) and conduction band minimum (CBM) with respect to the vacuum level, and the bandgap ($E_g$). 
       All energies are given in eV and the pressures is in kbar ($10^{8}$ Pa).
       \label{tbl:hkust}}
     \begin{ruledtabular}
     \begin{tabular}{d d d d d}
      \text{Iteration} & P & \text{VBM}   & \text{CBM}  & E_g \\ \hline
        0     &   -13.98     & -7.5  & -3.7 & 3.8 \\
        1     &   -1.09      & -7.0  & -3.5 & 3.5 \\
     \end{tabular}
     \end{ruledtabular}
   \end{center}
 \end{table}
 
 \section{Conclusions}
The rapid volume optimisation (RVO) approach presented here uses information from an inexpensive energy-volume curve to obtain a useful estimate of the optimal unit cell volume for a different level of theory. 
Our focus was in bridging between different exchange-correlation functionals within density functional theory, but a measured bulk modulus or classical interatomic potential could also be used to construct the reference energy-volume data.
In sensitive systems the volume change can lead to qualitative differences in the electronic and magnetic properties.
The results depend on the initial volume estimate and are relatively insensitive to the accuracy of the E-V curve.
The RVO method is expected to be competitive with conventional optimisation approaches for simple symmetric unit cells, as demonstrated for rocksalt structured PbS.
For materials such as \ce{Cu2ZnSnS4} that are sensitive to the atomic positions within the unit cell,
RVO may form part of the optimisation approach but direct optimisation of internal positions is still needed.
More significantly, it allows for the improved estimation of properties for large unit cells as demonstrated for HKUST-1, where conventional optimisation methods may be infeasible.
As the only property used is the hydrostatic pressure, it is possible to employ calculation methods which return a total energy without analytical gradients by evaluating the energy of a single finite difference.
In this case, an improvement over the ``single-shot'' may be obtained with two additional high-level calculations and an inexpensive E-V curve; a fourth high-level calculation would give an estimate of the convergence.
We expect that in many cases this will prove an economical approach for the application of state-of-the-art electronic structure calculations in the solid state.

\section{Supplementary data}
Python code providing a reference implementation of this method is available at \url{http://github.com/wmd-bath/rvo}; a snapshot as of this publication is available at DOI:10.5281/zenodo.31940.
This includes a program to generate the plots in this publication from the binary chalcogenide data.
Calculation data has been deposited online with Figshare at the DOI: 10.6084/m9.figshare.1468388. 
Raw output files from the hybrid DFT calculations are made available for CZTS and HKUST-1, and energy-volume curves are available for all the systems.
The full set of graphs and fitting parameters for PbS, PbTe, ZnS, and ZnTe are also included as supplementary data with this paper.

\acknowledgements
The authors thank J. M. Frost for useful discussions.
We acknowledge membership of the UK's HPC Materials Chemistry Consortium, which is funded by EPSRC grant EP/L000202.
J.M.S. is funded by an EPSRC Programme Grant (EP/K004956/1). 
A.J.J. is funded by the EPSRC Doctoral Training Centre in Sustainable Chemical Technologies (EP/G03768X/1 and  EP/L016354/1).
A.W. acknowledges support from the Royal Society and the ERC (grant no. 277757).

\bibliography{dvxc}

\end{document}